\documentclass[preprint,epsfig,floats,aps]{revtex4}
\usepackage{epsfig}
\begin{document}
\topmargin=-0.5cm
\title{Charged multiplicity density and number of\\ participant nucleons 
       in relativistic nuclear collisions}
\author{Ben-Hao Sa$^{1,4,5,6}$,
 A. Bonasera$^2$, Xu Cai$^{5,4}$, 
 An Tai$^3$, and Dai-Mei Zhou$^4$}
\affiliation{
$^1$  China Institute of Atomic Energy, P. O. Box 275 (18),
      Beijing, 102413 China \\
$^2$  Laboratorio Nazionale del Sud, Instituto Nazionale Di Fisica Nucleare,
      v. S. Sofia 44 95132 Catania, Italy \\
$^3$  Department of Physics and Astronomy, University of California,
      at Los Angeles, Los Angeles, CA 90095 USA \\
$^4$  Institute of Particle Physics, Huazhong Normal University,
      Wuhan, 430079 China\\
$^5$  CCAST (World Lab.), P. O. Box 8730 Beijing, 100080 China\\
$^6$  Institute of Theoretical Physics, Academia Sinica, Beijing,
      100080 China
\footnote{Email: sabh@iris.ciae.ac.cn; bonasera@lns.infn.it}
}
\begin{abstract}
The energy and centrality dependences of charged particle pseudorapidity 
density in relativistic nuclear collisions were studied using a hadron and 
string cascade model, JPCIAE. Both the relativistic $p+\bar p$ experimental 
data and the PHOBOS and PHENIX $Au+Au$ data at RHIC energy could be fairly 
reproduced within the framework of JPCIAE model and without retuning 
the model parameters. The predictions for $Pb+Pb$ collisions at the LHC  
energy were also given. We computed the participant nucleon distributions 
using different methods. It was found that the number of participant nucleons 
is not a well defined variable both experimentally and theoretically. 
Thus it may be inappropriate to use the charged particle pseudorapidity 
density per participant pair as a function of the number of participant 
nucleons for distinguishing various theoretical models. A discussion for the 
effect of different definitions in nuclear radius (diffused or sharp) was 
given.
\end{abstract}
\maketitle

\section{\bf{INTRODUCTION}}

The main focus of the Relativistic Heavy-Ion Collider (RHIC) at the Brookhaven 
National Laboratory (BNL) is to explore the phase transition related to 
the quark deconfinement and the chiral symmetry restoration. The first 
available experimental data were the energy dependence of charged particle 
pseudorapidity density in central $Au+Au$ collisions at $\sqrt s_{nn}$=56 and 
130 GeV from the PHOBOS collaboration \cite{pho1}. Soon later, the PHENIX 
collaboration published their data of centrality dependence of the charged 
particle pseudorapidity density in $Au+Au$ collisions at $\sqrt s_{nn}$=130 
GeV \cite{phe1}. Recently the charged particle pseudorapidity density in 
central $Au+Au$ collisions at $\sqrt s_{nn}$=200 GeV was also reported 
by PHOBOS collaboration \cite{pho3}. The BRAHMS collaboration published  
more recently their results of charged particle density from Au+Au collisions 
at $\sqrt s_{nn}$=130 GeV \cite{bra1,bra2}.

\begin{figure}[ht]
\centerline{\hspace{-0.5in}
\epsfig{file=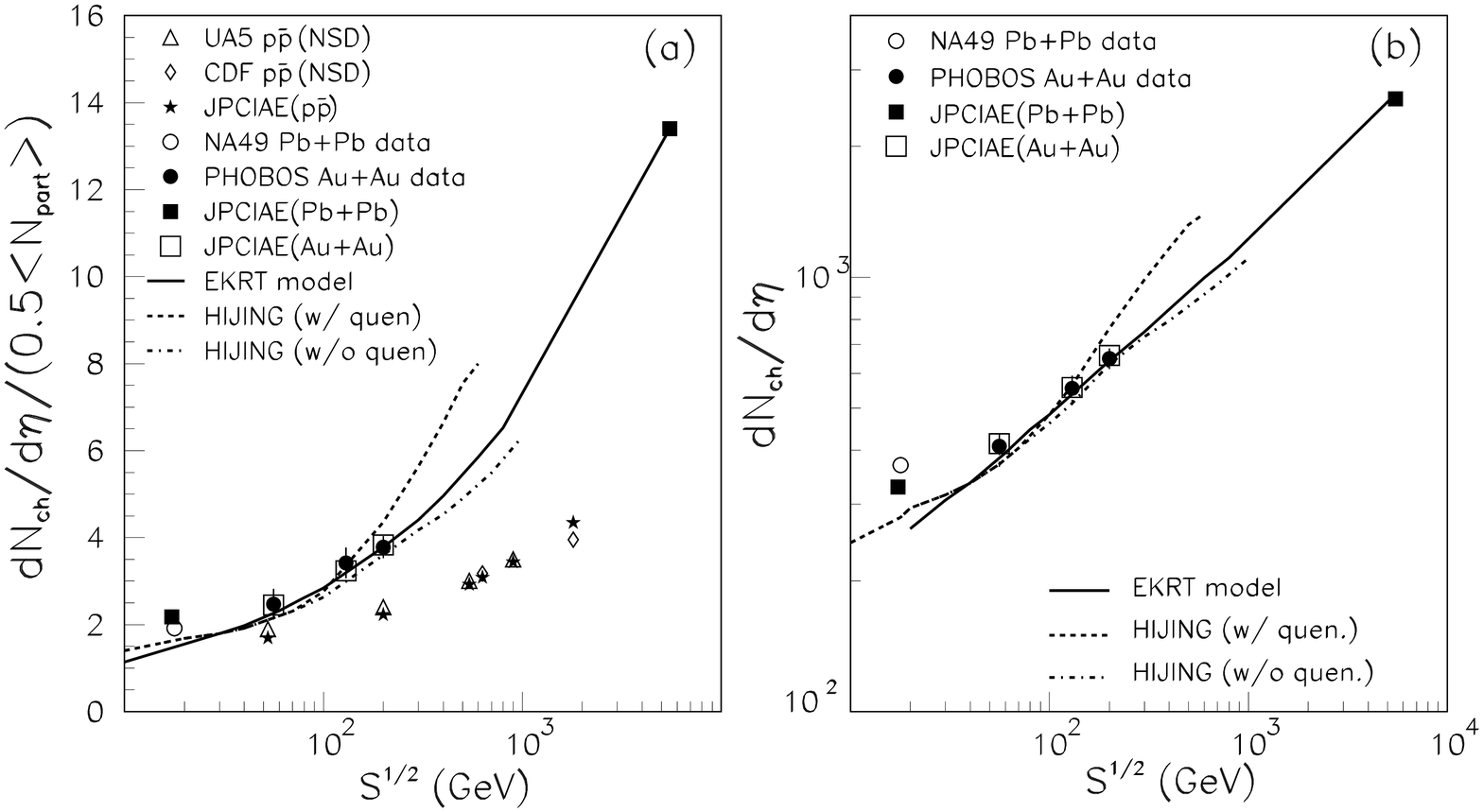,width=4.8in,height=3.15in,angle=0}}
\caption{The energy dependence of the charged particle pseudorapidity density 
at mid-pseudorapidity in relativistic $p+\bar p$ and central $A+A$ collisions.}
\label{fig1_p}
\end{figure}

It has been predicted that the rare high charged multiplicity in the final 
state of relativistic nucleus-nucleus collisions might indicate the formation 
of the Quark-Gluon-Plasma (QGP) phase in the early stage of collisions 
\cite{hov,gor,kap1}. In Ref. \cite{wang1}, the centrality dependence of the 
charged multiplicity has been further proposed to provide information on the 
relative importance of soft versus hard processes in particle production and 
therefore to provide a means of distinguishing various theoretical models. 

 The pQCD calculation with assumption of gluon saturation \cite{esk1,esk2} 
(referred to as the EKRT model later) was first used to study the centrality 
dependence of the charged particle pseudorapidity density at RHIC. The 
conventional eikonal approach and the high density QCD (referred to as the KN 
model later) \cite{kha1} were also used to investigate the centrality 
dependence and these two methods surprisingly obtained almost identical 
centrality dependence. Recently, authors in \cite{cape} reported their study
of the same issue from the Dual Parton Model. It was found that the 
experimental observation, the charged particle pseudorapidity density per 
participant pair slightly increasing with the number of participant nucleons, 
was reproduced by \cite{wang1,kha1,cape}, but contradicted the results of 
\cite{esk2}.   

\begin{figure}[ht]
\centerline{\hspace{-0.5in}
\epsfig{file=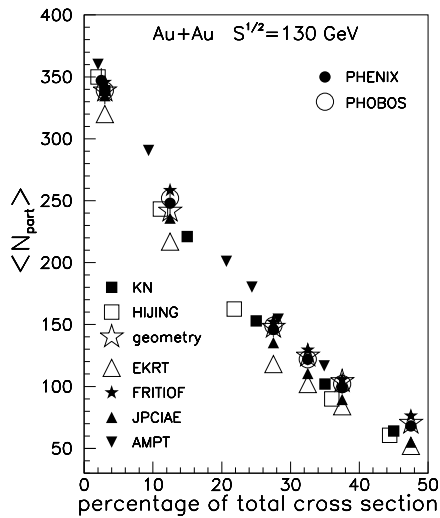,width=3.0in,height=3.0in,angle=0}}
\caption{The number of participant nucleons $<N_{part}>$ as a 
function of the percentage of total cross section.}
\label{fig2_p}
\end{figure}

In this paper a hadron and string cascade model, JPCIAE \cite{sa1}, was
employed to study this issue further. Within the framework of this model
the experimentally measured energy dependence of the charged particle
mid-pseudorapidity density per participant pair both in relativistic $p+\bar p$
and $Au+Au$ collisions at RHIC energies was reproduced fairly well without
retuning the model parameters. The predictions for the $Pb+Pb$ collisions at 
LHC energy were also given. Both the PHENIX \cite{phe1} and the PHOBOS 
\cite{pho2} observations that the charged particle mid-pseudorapidity density 
per participant pair slightly increases with the number of participant 
nucleons could be reproduced fairly well by JPCIAE. In studying centrality 
dependence the focus was put on the uncertainties in the definitions and 
calculations of the number of participant nucleons. However, this study 
indicated that it is not suitable to use the charged particle mid-
pseudorapidity density per participant pair as a function of the number of 
participant nucleons to constrain theoretical models for particle production, 
because the number of participant nucleons is not a well defined physical 
variable both experimentally and theoretically. The effect of the different 
definitions in nuclear radius (diffused or sharp) on the number of 
participant nucleons, the charged particle mid-pseudorapidity density, and 
the charged particle mid-pseudorapidity density per participant pair as a 
function of the number of participant nucleons was discussed. A brief version 
of the part contents of this paper has been published elsewhere \cite{sa2}.

\begin{figure}[ht]
\centerline{\hspace{-0.5in}
\epsfig{file=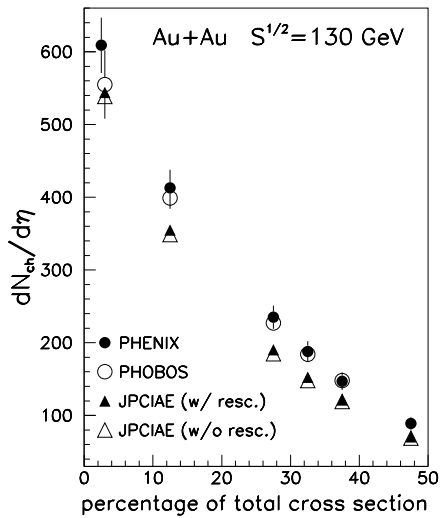,width=3.0in,height=3.0in,angle=0}}
\caption{The charged particle pseudorapidity density at mid-pseudorapidity 
 in $Au+Au$ collisions at $\sqrt s_{nn}$=130 GeV as a function of the 
 percentage of total cross section.}
\label{fig3_p}
\end{figure}

\section{\bf{MODELS}}

The JPCIAE model was developed based on PYTHIA \cite{sjo1}, which is a well 
known event generator for hadron-hadron collisions. In the JPCIAE model the
radial position of a nucleon in colliding nucleus $A$ (indicating the atomic
number of this nucleus as well) is sampled randomly according to the
Woods-Saxon distribution
 \begin{equation}
 \rho(r) \sim \frac{\rho_0}{1+exp((r-R_A)/\alpha)}
 \end{equation} 
where $\rho_0$ refers to the normal nuclear density, $\alpha \sim$ 0.54 fm 
stands for the nuclear diffusion edge, and $R_A$ is the nuclear radius of 
nucleus $A$. The solid angle of the nucleon is sampled
uniformly in 4$\pi$. Each nucleon is given a beam momentum in z direction and
zero initial momenta in x and y directions. The collision time of each
colliding pair is calculated under the requirement that the least approach
distance of the colliding pair along their straight line trajectory
(mean field potential is not taken into account in JPCIAE) should be
smaller than $\displaystyle{\sqrt{\sigma_{tot}/\pi}}$. Here $\sigma_{tot}$
refers to the total cross section. The nucleon-nucleon collision with the
least collision time is then selected from the initial collision list to
perform the first collision. Both the particle list and the collision list are
then updated such that the new collision list may consist of not only nucleon-
nucleon collisions but also collisions between nucleons and produced
particles and between produced particles themselves. The next collision is
selected from the new collision list and the processes above are repeated 
until the collision list is empty.
   
\begin{figure}[ht]
\centerline{\hspace{-0.5in}
\epsfig{file=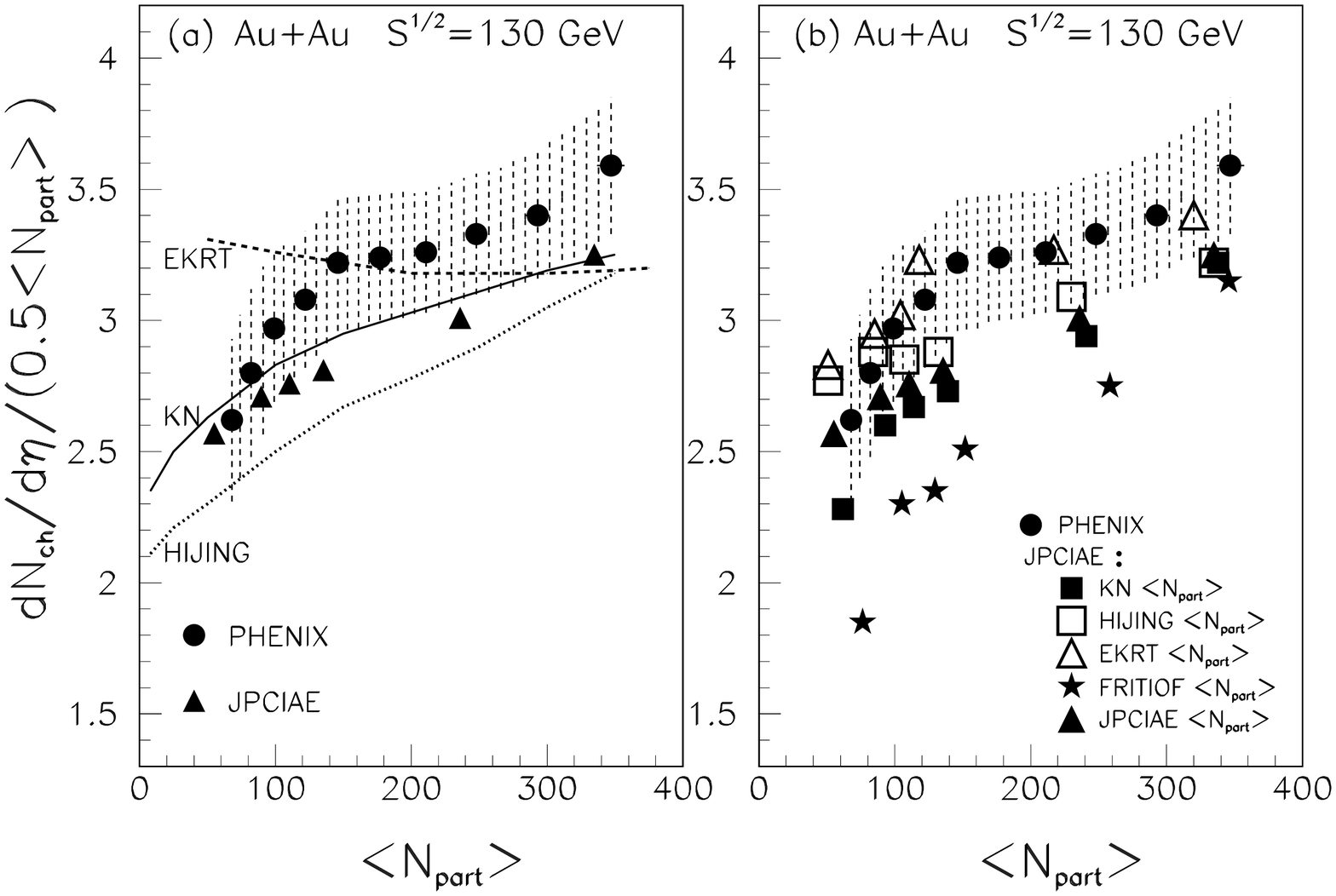,width=4.8in,height=3.15in,angle=0}}
\caption{The charged particle pseudorapidity density per participant pair  
 at mid-pseudorapidity in $Au+Au$ collisions at $\sqrt s_{nn}$=130 GeV as 
 a function of the number of participant nucleons, $<N_{part}>$.}
\label{fig4_p}
\end{figure}

 For each collision pair, if its CMS energy is larger than a given cut,
we assume that strings are formed after the collision and PYTHIA is used
to deal with particle production. Otherwise, the collision is treated as a
two-body collision \cite{cugn,bert,tai1}. The cut (=4 GeV in the 
program) was chosen by observing that JPCIAE correctly reproduces charged 
multiplicity distributions in $AA$ collisions \cite{sa1}. An important feature  
of JPCIAE at relativistic energies is that QCD parton-parton scatterings 
are included through PYTHIA \cite{marie}, which causes charged particle 
yields to increase with collision energy as well as centrality since mini-jet 
production rates increase with energy and number of collisions suffered by 
participant nucleons. It should be noted that the JPCIAE model is not a simple 
superposition of nucleon-nucleon collisions since the rescattering among 
participant nucleons, spectator nucleons, and produced particles is taken into 
account. We refer to \cite{sa1} for more about the JPCIAE model.    

 Since the number of participant nucleons, $N_{part}$, plays a crucial role 
in the presentation of PHOBOS or PHENIX data we first make a study on 
$N_{part}$. As the direct measurement of $N_{part}$ is not available, in the 
fixed target experiments the number of participant nucleons from the  
projectile nucleus with atomic number $A$, for instance, is estimated by
  \begin{equation}
  N_{part}^p=A*(1-\frac{E_{ZDC}}{E_{beam}^{kin}}),
  \end{equation}
where $E_{ZDC}$ refers to the energy deposited in the Zero Degree Calorimeter
(ZDC), dominated by the energy deposition from the projectile spectator 
nucleons, and $E_{beam}^{kin}$ is the kinetic energy of beam \cite{ahle}. 
However, in the collider experiments, in order to obtain $N_{part}$ one has to 
relate the measurements to the Monte Carlo simulations. In PHENIX, for 
instance, simulations for the response of the Beam-Beam Counter and the ZDC 
were used to calculate $N_{part}$ via the Glauber model \cite{phe1}. In 
PHOBOS, $N_{part}$ is derived by relating HIJING simulations to the signals 
in the paddle counter \cite{pho2}. Therefore, $N_{part}$ here is a model-
dependent variable.

\begin{figure}[ht]
\centerline{\hspace{-0.5in}
\epsfig{file=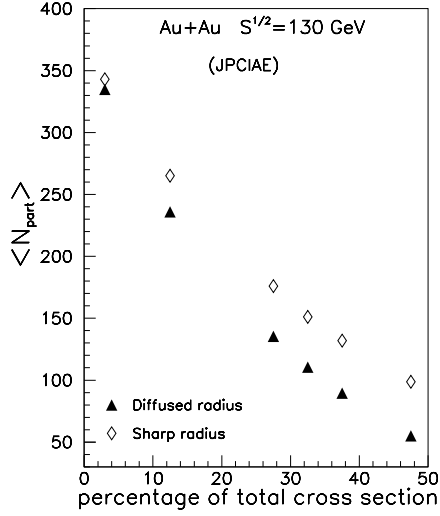,width=3.0in,height=3.0in,angle=0}}
\caption{The number of participant nucleons $<N_{part}>$ as a
function of the percentage of total cross section, triangles: diffused 
nuclear radius, rhombuses: sharp nuclear radius.}
\label{fig5_p}
\end{figure}

 Theoretically, the number of participant nucleons in a collision of $A+B$ 
at impact parameter $b$ can be estimated in different ways:
\begin{enumerate} 
\item In the geometry method \cite{sa0} $N_{part}(b)$ reads
 \begin{eqnarray}
  N_{part}(b)=N_{part}^A(b)+N_{part}^B(b),\\
  N_{part}^A(b)=\rho_A\displaystyle{\int} {dV\theta({R_A-(x^2+(b-y)^2+z^2)
                 ^{1/2}}})\theta({R_B-(x^2+y^2)^{1/2}}),\\
  N_{part}^B(b)=\rho_B\displaystyle{\int} {dV\theta({R_B-(x^2+y^2+z^2)
                 ^{1/2}}})\theta({R_A-(x^2+(b-y)^2)^{1/2}}),  
 \end{eqnarray}
where $\theta(x)$=0 if $x<$0 and $\theta(x)$=1 otherwise, $\rho_A$ ($\rho_B$) 
is the nuclear density of the projectile (target) nucleus and the nuclear 
density is normalized to the atomic number. 
\item In the Glauber model,  $N_{part}(b)$ is calculated through    
 \begin{equation}
 N_{part}(b)=\int{d^2sT_A(\vec b-\vec s)[1-\exp(-\sigma_{in}T_B(\vec s))]}
              +\int{d^2sT_B(\vec s)[1-\exp(-\sigma_{in}T_A(\vec b-\vec s))]}
 \end{equation}
where $\sigma_{in}\approx$40 mb is the inelastic nucleon-nucleon cross section 
at RHIC energies and $T_A$ ($T_B$) refers to the nuclear thickness function of 
nucleus $A$ ($B$) and is normalized to $A$ ($B$) \cite{esk2}. 
\item In the dynamical simulation of $A+B$ collisions, $N_{part}(b)$ can 
be estimated via counting the participant or spectator nucleons and averaging 
over events simulated at a given impact parameter $b$. However, there 
are multifarious in simulating models such as: FRITIOF \cite{pi1}, 
VENUS \cite{venu}, HIJING \cite{wang2}, JPCIAE \cite{sa1}, UrQMD \cite{uqmd},  
and AMPT \cite{ampt}, etc. . Not only is the theoretical uncertainty related to
the calculation of $N_{part}(b)$ in each of the above models large but also 
the definitions of participant nucleons or spectator nucleons are different 
among each other. We give only a necessary description for the following 
dynamical simulations mentioned in this paper:
 \begin{itemize}   
 \item In FRITIOF \cite{pi1} and HIJING \cite{wang2}, the wounded nucleons, 
i.e., nucleons which suffer at least one inelastic collision, are counted and 
identified as $N_{part}(b)$. It should be pointed out here that in FRITIOF and 
HIJING the produced particles from the string fragmentation do not have 
rescattering.  
 \item In JPCIAE, we have counted the nucleons involved in at least one 
inelastic nucleon-nucleon collision with string excitation and identified 
them as $N_{part}(b)$.
 \item In AMPT \cite{ampt1}, the spectator nucleons, $N_{spec}(b)$, are counted 
in the final state without rescattering and $N_{part}(b)$ is then calculated 
through
 \begin{equation}
 N_{part}(b)=(A+B)-N_{spec}(b).
 \end{equation}   
The spectator nucleons here refer to the nucleons with zero transverse 
momentum and beam energy in the final state of the AMPT simulation.  
 \end{itemize}
\end{enumerate}

\begin{figure}[ht]
\centerline{\hspace{-0.5in}
\epsfig{file=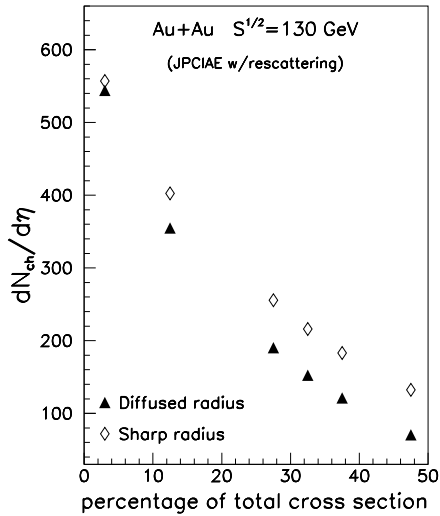,width=3.0in,height=3.0in,angle=0}}
\caption{The charged particle pseudorapidity density at mid-pseudorapidity
 in $Au+Au$ collisions at $\sqrt s_{nn}$=130 GeV as a function of the
 percentage of total cross section, triangles: diffused nuclear radius, 
 rhombuses: sharp nuclear radius.}
\label{fig6_p}
\end{figure}

 In the PHENIX or PHOBOS experiment, centrality bin was defined by the cut 
in the particle multiplicity distribution and expressed as the percentage of 
geometrical (total) cross section \cite{phe1}, $g$. However, in the 
theoretical calculation it is more convenient to define the centrality by 
impact parameter $b$. In order to compare the experimental data of centrality 
dependence with the theoretical results a relation between $g$ and $b$ is 
required, which is obtained from the definition of the geometrical cross 
section, that reads     
 \begin{equation}
 b=\sqrt g \times b_{max}, b_{max}=R_A+R_B, 
 \end{equation}  
where $R_A=1.12A^{1/3}+0.54$ fm, for instance, is the radius of nucleus $A$ 
with diffused edge. However, $R_{Au}$= 7.5 fm is taken in our calculations 
since this value (corresponding to $\sigma_{geo.}$=7.2 b) was used in 
\cite{phe1} to extract the number of participant nucleons in $Au+Au$ 
collisions. We give in Tab. 1 the mapping between the centrality bin 
($g$ bin, selected from Tab. 1 in \cite{phe1}) and the $b$ bin according to 
Eq. (8), and the averaged impact parameter, $\bar b$, over the $b$ bin 
according to the $b^2$ law.   

\begin{table}[hb]
\begin{tabular}{c|c|c}
\multicolumn{3}{c}{Table 1. The mapping between the $g$ bin}\\ 
\multicolumn{3}{c}{and the $b$ bin in $Au+Au$ collisions}\\
\multicolumn{3}{c}{(case of diffused nuclear radius)}\\  
\hline\hline
 centrality bin &bin of $b$ (fm) &$\bar b$ (fm)\\
\hline
 below 6$^*$ &below 3.67 &2.45 \\
 10 - 15 &4.74 - 5.81 &5.29 \\
 25 - 30 &7.50 - 8.22 &7.87 \\
 30 - 35 &8.22 - 8.87 &8.55 \\
 35 - 40 &8.87 - 9.49 &9.18 \\
 45 - 50 &10.1 - 10.6 &10.4 \\
\hline
\hline
\multicolumn{3}{c}{$^*$ below 5 for PHENIX.}
\end{tabular}
\end{table}

As the experimental data were averaged over events in each $g$ bin the 
theoretical results, to be compared with the experimental data, are also 
averaged over the corresponding $b$ bin. We denote the number of 
participant nucleons after averaging over $g$ or $b$ bin as $<N_{part}>$ 
later. 

\section{\bf{RESULTS AND DISCUSSIONS}}

 In Fig. 1(a) the experimental data of charged particle pseudorapidity 
density per participant pair at mid-pseudorapidity in relativistic $p+\bar p$ 
(open triangles and rhombuses with error bar) and in central $A+A$ collisions 
(open circles with error bar for $Pb+Pb$ at SPS and full circles with error 
bar for $Au+Au$ at RHIC) \cite{pho1,pho3} were compared with the results of 
the JPCIAE model (full stars for $p+\bar p$, open squares for $Au+Au$ 
collisions at RHIC, and full squares for $Pb+Pb$ 
at $\sqrt s_{nn}$=17.3 and 5500 GeV). In addition, the results from other 
models were plotted as follows: the dashed and dotted-dash curves are from  
the HIJING model \cite{wang1} with and without jet quenching, respectively, 
the solid curve are from the EKRT model \cite{esk2}. The EKRT results   
were obtained from \cite{wang1} directly, except that the EKRT result for 
$Au+Au$ collisions at $\sqrt s_{nn}$=5500 GeV was taken from \cite{esk3}. 
Fig. 1 (b) is the same as (a) but for $A+A$ collisions only and the vertical 
coordinate here is the charged particle pseudorapidity density itself. One 
knows from Fig. 1 that both the data of $p+\bar p$ and $A+A$ collisions at 
relativistic energies are also reproduced fairly well by the JPCIAE model. 

\begin{figure}[ht]
\centerline{\hspace{-0.5in}
\epsfig{file=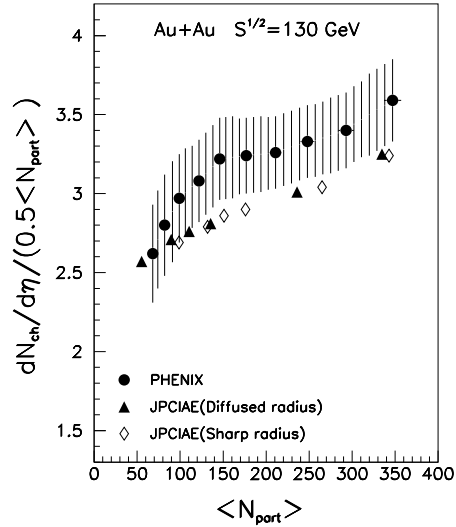,width=3.0in,height=3.0in,angle=0}}
\caption{The charged particle pseudorapidity density per participant pair
 at mid-pseudorapidity in $Au+Au$ collisions at $\sqrt s_{nn}$=130 GeV as
 a function of $<N_{part}>$. The circles with shaded area of systematic 
 errors: PHENIX data, triangles: JPCIAE (diffused radius), rhombuses: JPCIAE
 (sharp radius).}
\label{fig7_p}
\end{figure}

 In Fig. 2 the $<N_{part}>$ extracted by PHENIX \cite{phe1} and PHOBOS 
\cite{pho2} from $Au+Au$ collisions at $\sqrt s_{nn}$=130 GeV were compared 
with the model computations. The horizontal axis in Fig. 2 is the percentage 
of geometrical cross section, $g$, and each $g$ bin is represented by its 
middle point for convenience in plotting (the same for Fig. 3, 5, and 6). The 
solid and open circles with error bar in Fig. 2 were PHENIX and PHOBOS 
results, respectively. $<N_{part}>$ from the geometry method, open stars, 
were from the geometric $N_{part}(b)$ (Eq. 3 - 5) after averaging over $b$ 
sampled randomly in each $b$ bin due to the $b^2$ law. Similarly, the FRITIOF, 
JPCIAE, and AMPT \cite{ampt1} results, full stars, full triangle-ups, and full 
triangle-downs in Fig. 2, were obtained averaging over events simulated for 
$b$ sampled randomly in each $b$ bin due to the $b^2$ law, respectively. 
In the EKRT model \cite{esk2}, the curves of $N_{part}(b)$ vs. $b/R_A$ were 
given and $N_{part}$ is calculated by the Glauber method with $R_A$=1.12$A^
{1/3}$-0.86$A^{-1/3}$. The open triangle-ups in Fig. 2 are the EKRT model 
results taken from the $\sqrt s_{nn}$=130 GeV one of those curves according to 
the $\bar b$ in Tab. 1. Originally, HIJING $N_{part}$ \cite{wang3} were 
calculated for individual $b$, the HIJING points, open squares, in Fig. 2 were 
plotted after relating $b$ to $g$ according to Eq. (8). The full squares in 
Fig. 2 were the results of the KN model taken from Tab. 1 in \cite{kha1} under 
the centrality bins of 0 - 6, 10 - 20, 20 - 30, 30 - 40, and 40 - 50 \%, 
respectively. In the KN model, $<N_{part}>$ was the result of $N_{part}(b)$ 
from the Glauber model after averaging over centrality bin. In this approach, 
the particle multiplicity and impact parameter were related by a Gaussian 
distribution with parameters fixed via fitting the PHOBOS charged multiplicity 
distribution. As proved in \cite{wojc}, such kind of average is approximately 
equivalent to the average method based on Eq.(8). From Fig. 2 one knows that 
the discrepancies among the PHENIX or PHOBOS and the model results are 
visible. 
       
 The charged particle pseudorapidity density at mid-pseudorapidity in $Au+Au$ 
collisions at $\sqrt s_{nn}$=130 GeV as a function of the percentage of 
geometrical cross section was given in Fig. 3. In this figure, the full and 
open circles with error bar are the PHENIX \cite{phe1} and PHOBOS \cite{pho2} 
data, respectively. The full and open triangles, respectively, are the JPCIAE  
results with and without rescattering. One knows from Fig. 3 that the 
rescattering only leads to a few percent increase in the charged multiplicity 
although rescattering might enhance yields of strangeness, $\Xi^- + \overline
{\Xi^-}$ for instance, by a couple of times.  

\begin{table}[hb]
\begin{tabular}{c|c|c}
\multicolumn{3}{c}{Table 2. The mapping between the $g$ bin}\\
\multicolumn{3}{c}{and the $b$ bin in $Au+Au$ collisions}\\
\multicolumn{3}{c}{(case of sharp nuclear radius)}\\
\hline\hline
 centrality bin &bin of $b$ (fm) &$\bar b$ (fm)\\
\hline
 below 6$^*$ &below 3.20 &2.13 \\
 10 - 15 &4.11 - 5.03 &4.59 \\
 25 - 30 &6.50 - 7.12 &6.82 \\
 30 - 35 &7.12 - 7.69 &7.41 \\
 35 - 40 &7.69 - 8.22 &7.96 \\
 45 - 50 &8.72 - 9.19 &8.86 \\
\hline
\hline
\multicolumn{3}{c}{$^*$ below 5 for PHENIX.}
\end{tabular}
\end{table}

 In Fig. 4 (a) we compared the PHENIX data of charged particle 
mid-pseudorapidity density per participant pair (full circles with shaded area 
of systematic errors) \cite{phe1} with the results of the JPCIAE model (full 
triangles) and the results of other models (obtained from \cite{phe1} 
directly): HIJING (the dotted curve), the KN model (the solid curve), and 
EKRT (the dashed curve). One sees that except EKRT, three other models predict 
an increase of $\displaystyle (dN_{ch}/d\eta|_{\eta=0})/(0.5<N_{part}>)$ as a 
function of $<N_{part}>$ though the theoretical results seem to underestimate 
the PHENIX data.  Such an increase can be understood in JPCIAE as a result of 
increasing hard parton scatterings per participant nucleon. Fig. 4 (b) 
compared the PHENIX data to the results of single $dN_{ch}/d\eta|_{\eta=0}$ 
from JPCIAE normalized by the $<N_{part}>$ from different models (taken from 
the corresponding curve in Fig. 2 at the middle point of $g$ bins for KN, 
HIJING, and EKRT models): full squares, $<N_{part}>$ from the KN model, open 
squares from HIJING, open triangles from EKRT, full stars from FRITIOF, and 
full triangles from JPCIAE. One sees from Fig. 4 (b) that starting from the 
charged particle mid-pseudorapidity density obtained in JPCIAE, but using 
$<N_{part}>$ from different models, the corresponding results of 
$\displaystyle (dN_{ch}/d\eta|_{\eta=0})/(0.5<N_{part}>)$ are different 
visibly among each other, in peripheral collisions especially. Therefore it 
might be inappropriate using $\displaystyle (dN_{ch}/d \eta|_{\eta=0})/
(0.5<N_{part}>)$ as a function of $<N_{part}>$ to distinguish various 
theoretical models for particle production since $<N_{part}>$ is not a well 
defined physical variable.  

If one uses the sharp nuclear radius, $R_A=1.12A^{1/3}$ fm ($R_{Au}$=6.50 fm), 
instead of diffused one used above, the corresponding mapping between the 
centrality bin ($g$ bin) and the $b$ bin from Eq. (8) is given in Tab. 2 . 
Fig. 5, 6, and 7 give, respectively, the comparisons between the JPCIAE 
results of diffused nuclear radius and the JPCIAE results of sharp nuclear 
radius in the $<N_{part}>$ as a function of centrality, the 
$dN_{ch}/d\eta|_{\eta=0}$ as a function of centrality, and the
$(dN_{ch}/d\eta|_{\eta=0})/(0.5<N_{part}>)$ as a function of $<N_{part}>$.
One knows from Fig. 5 and 6 that there are observable discrepancies between 
the results of diffused nuclear radius and the results of sharp nuclear radius 
, in peripheral collisions especially. However, in the  
$(dN_{ch}/d\eta|_{\eta=0})/(0.5<N_{part}>)$ vs. $<N_{part}>$ plot both results 
are close to each other as shown in Fig. 7 . That indicates again that it  
is hard using $\displaystyle (dN_{ch}/d \eta|_{\eta=0})/(0.5<N_{part}>)$ as a 
function of $<N_{part}>$ to distinguish various theoretical models for 
particle production since $N_{part}$ is not a well defined variable.
As one sees from Fig. 5 and 6 that for most central collision both diffused 
and sharp nuclear radii work well in mapping centrality bin and $b$ bin 
therefore we do not redraw a figure as Fig. 1 for sharp nuclear radius. 
However, it should be pointed out that using sharp nuclear radius in mapping 
centrality bin and $b$ bin is inconsistent with that the nuclear diffusion 
edge is introduced in Woods-Saxon distribution initiating the nucleons in 
nucleus. 

\section{\bf{CONCLUSIONS}}

In summary, we used the hadron and string cascade model, JPCIAE, to investigate 
the energy and centrality dependences of charged particle pseudorapidity 
density at mid-pseudorapidity in relativistic $p+\bar p$ and $A+A$ collisions. 
Both the relativistic $p+\bar p$ experimental data and the PHOBOS and PHENIX 
data of $Au+Au$ collisions at RHIC energies could be reproduced fairly well 
within the framework of the JPCIAE model without retuning any parameters. The 
JPCIAE model predictions for $Pb+Pb$ collisions at the LHC energy were also 
given. Both the PHENIX \cite{phe1} and the PHOBOS \cite{pho2} observations 
that the charged particle mid-pseudorapidity density per participant pair 
slightly increases with the number of participant nucleons could be 
reproduced fairly well by JPCIAE. This study shows that since $<N_{part}>$ is 
not a well defined physical variable both experimentally and theoretically it 
may be hard to use charged particle pseudorapidity density per participant 
pair at mid-pseudorapidity as a function of $<N_{part}>$ to distinguish 
various theoretical models for particle production. A discussion for the 
effects of the different definitions in nuclear radius (diffused or sharp) is 
given and it is indicatd that using diffused nuclear radius might be better.

{\bf{ACKNOWLEDGMENTS}}
              
Finally, the financial supports from NSFC (19975075, 10135030, and 10075035) 
in China and DOE in USA are acknowledged.

\end{document}